\documentclass[12pt]{article}
\usepackage[dvips]{epsfig}
\begin{document}                                                                
\date{}

\title{
{\vspace{-1em} \normalsize                                                      
\hfill \parbox{50mm}{DESY 99-170}}\\[25mm]
{\Large\bf Least-squares optimized polynomials for           \\  
       fermion simulations}\footnote{Talk given at the workshop
       on {\em Numerical Challenges in Lattice QCD},
       August 1999, Wuppertal University; 
       to appear in the proceedings.}}

\author{I. Montvay                                           \\[0.5em]
        Deutsches Elektronen-Synchrotron DESY,               \\
        Notkestr.\,85, D-22603 Hamburg, Germany              \\
        E-mail: istvan.montvay@desy.de}

\newcommand{\half}{\frac{1}{2}}                                                 
\newcommand{\rar}{\rightarrow}                                                  
\newcommand{\lar}{\leftarrow}

\maketitle

\begin{abstract}
 Least-squares optimized polynomials are discussed which are needed in
 the two-step multi-bosonic algorithm for Monte Carlo simulations of
 quantum field theories with fermions.
 A recurrence scheme for the calculation of necessary coefficients in
 the recursion and for the evaluation of these polynomials is
 introduced.
\end{abstract}

\section{Introduction}

 In popular Monte Carlo simulation algorithms for QCD and other similar
 quantum field theories the main difficulty is the evaluation of the
 determinant of the fermion action matrix.
 This can be achieved by stochastic procedures with the help of
 auxiliary bosonic ``pseudofermion'' fields.
 
 In the {\em two-step multi-bosonic (TSMB) algorithm}~\cite{TWO-STEP}
 an approximation of the fermion determinant is achieved by the
 pseudofermion fields corresponding to a  polynomial approximation of
 some negative power $x^{-\alpha}$ of the fermion matrix~\cite{POLYNOM}.
 The auxiliary bosonic fields are updated according to the
 {\em multi-bosonic action}~\cite{MULTIBOSON}.
 The error of the polynomial approximation is corrected in a global
 accept-reject decision by using better polynomial approximations.
 Sometimes a reweighting of gauge configurations in the evaluation
 of expectation values is also necessary.
 This can also be performed by high order polynomials.

 The polynomials used in the TSMB algorithm have to approximate the
 function $x^{-\alpha}\bar{P}(x)$ in some non-negative interval
 $x \in [\epsilon,\lambda],\; 0 \leq \epsilon < \lambda$.
 Here $\bar{P}(x)$ is a known polynomial, typically another cruder
 approximation of $x^{-\alpha}$.
 The approximation scheme and optimization procedure can be chosen
 differently.
 The least-squares optimization \cite{APPROXIMA} is an efficient and
 flexible possibility.
 (For other approximation schemes see \cite{SCHRODING,UVFILTER}.)

 In this review the basic relations for least-squares optimized
 polynomials are presented as introduced in \cite{TWO-STEP,POLYNOM}.
 Particular attention is paid to a recurrence scheme which can be
 applied for determining the necessary high order polynomials and
 for evaluating them numerically.
 The details of the TSMB algorithm will not be considered.
 For a comprehensive summary and references see \cite{MOLDYN}.
 For experience on the application of TSMB in a recent large scale
 numerical simulation see \cite{DESYMUN1,DESYMUN2}.

\section{Basic relations}

 Least-squares optimization provides a general and flexible framework
 for obtaining the necessary optimized polynomials in multi-bosonic
 fermion algorithms.
 Here we introduce the basic formulae in the way it has been done in
 \cite{POLYNOM,DESYMUN2}.

 We want to approximate the real function $f(x)$ in the interval
 $x \in [\epsilon,\lambda]$ by a polynomial $P_n(x)$ of degree $n$.
 The aim is to minimize the deviation norm
\begin{equation} \label{eq_01}
\delta_n \equiv \left\{ N_{\epsilon,\lambda}^{-1}\;
\int_\epsilon^\lambda dx\, w(x)^2 \left[ f(x) - P_n(x) \right]^2
\right\}^\half \ .
\end{equation}
 Here $w(x)$ is an arbitrary real weight function and the overall
 normalization factor $N_{\epsilon,\lambda}$ can be chosen by
 convenience, for instance, as
\begin{equation} \label{eq_02}
N_{\epsilon,\lambda} \equiv
\int_\epsilon^\lambda dx\, w(x)^2 f(x)^2 \ .
\end{equation}
 A typical example of functions to be approximated is
 $f(x)=x^{-\alpha}/\bar{P}(x)$ with $\alpha > 0$ and some polynomial
 $\bar{P}(x)$.
 The interval is usually such that $0 \leq \epsilon < \lambda$.
 For optimizing the relative deviation one takes a weight function
 $w(x) = f(x)^{-1}$.

 $\delta_n^2$ is a quadratic form in the coefficients of the polynomial
 which can be straightforwardly minimized.
 Let us now consider, for simplicity, only the relative deviation
 from the simple function $f(x)=x^{-\alpha}=w(x)^{-1}$.
 Let us denote the polynomial corresponding to the minimum of $\delta_n$
 by
\begin{equation} \label{eq_03}
P_n(\alpha;\epsilon,\lambda;x) \equiv
\sum_{\nu=0}^n c_{n\nu}(\alpha;\epsilon,\lambda) x^{n-\nu} \ .
\end{equation} 
 Performing the integral in $\delta_n^2$ term by term we obtain
\begin{equation} \label{eq_04}
\delta_n^2 = 1 -2 \sum_{\nu=0}^n c_\nu V_\nu^{(\alpha)}
+ \sum_{\nu_1,\nu_2=0}^n c_{\nu_1} M_{\nu_1,\nu_2}^{(\alpha)} c_{\nu_2}
\ ,
\end{equation} 
 where
$$
V_\nu^{(\alpha)} =
\frac{\lambda^{1+\alpha+n-\nu}-\epsilon^{1+\alpha+n-\nu}}
              {(\lambda - \epsilon)(1+\alpha+n-\nu)} \ ,
$$
\begin{equation} \label{eq_05} 
M_{\nu_1,\nu_2}^{(\alpha)} = 
\frac{\lambda^{1+2\alpha+2n-\nu_1-\nu_2} -
     \epsilon^{1+2\alpha+2n-\nu_1-\nu_2}}
              {(\lambda - \epsilon)(1+2\alpha+2n-\nu_1-\nu_2)}  \ .
\end{equation} 

 The coefficients of the polynomial corresponding to the minimum of
 $\delta_n^2$, or of $\delta_n$, are
\begin{equation} \label{eq_06}
c_\nu \equiv c_{n\nu}(\alpha;\epsilon,\lambda) =
\sum_{\nu_1=0}^n M_{\nu\nu_1}^{(\alpha)-1} V_{\nu_1}^{(\alpha)} \ . 
\end{equation} 
 The value at the minimum is
\begin{equation} \label{eq_07}
\delta_n^2 \equiv \delta_n^2(\alpha;\epsilon,\lambda) 
= 1 - \sum_{\nu_1,\nu_2=0}^n 
V_{\nu_1}^{(\alpha)} M_{\nu_1,\nu_2}^{(\alpha)-1} V_{\nu_2}^{(\alpha)}
\ .
\end{equation} 

 The solution of the quadratic optimization in 
 (\ref{eq_06})-(\ref{eq_07}) gives in principle a simple way to find
 the required least-squares optimized polynomials.
 The practical problem is, however, that the matrix $M^{(\alpha)}$ is
 not well conditioned because it has eigenvalues with very different
 magnitudes.
 In order to illustrate this let us consider the special case
 $(\alpha=1,\;\lambda=1,\;\epsilon=0)$ with $n=10$.
 In this case the eigenvalues are:
\begin{eqnarray} \label{eq_08}
& & 0.4435021205e-14 \ ,
\hspace{0.5em} 0.1045947635e-11 \ ,
\hspace{0.5em} 0.1143819915e-9  \ ,
\nonumber \\[0.5em] 
& & 0.7698917100e-8 \ ,
\hspace{1.0em} 0.3571195735e-6 \ ,
\hspace{1.0em} 0.1211873623e-4 \ ,
\nonumber \\[0.5em]
& & 0.3120413130e-3 \ ,
\hspace{1.0em} 0.6249495675e-2 \ ,
\hspace{1.0em} 0.9849331094e-1 \ ,
\nonumber \\[0.5em]
& & 1.075807246  \ .
\end{eqnarray} 
 A numerical investigation shows that, in general, the ratio of maximal
 to minimal eigenvalues is of the order of ${\cal O}(10^{1.5n})$.
 It is obvious from the structure of $M^{(\alpha)}$ in (\ref{eq_05})
 that a rescaling of the interval $[\epsilon,\lambda]$ does not
 help.
 The large differences in magnitude of the eigenvalues implies
 through (\ref{eq_06}) large differences of magnitude in the 
 coefficients $c_{n\nu}$ and therefore the numerical evaluation of
 the optimal polynomial $P_n(x)$ for large $n$ is non-trivial.

 Let us now return to the general case with arbitrary function $f(x)$
 and weight $w(x)$.
 It is very useful to introduce orthogonal polynomials
 $\Phi_\mu(x)\; (\mu=0,1,2,\ldots)$ satisfying
\begin{equation} \label{eq_09}
\int_\epsilon^\lambda dx\, w(x)^2 \Phi_\mu(x)\Phi_\nu(x)
= \delta_{\mu\nu} q_\nu \ .
\end{equation}
 and expand the polynomial $P_n(x)$ in terms of them:
\begin{equation} \label{eq_10}
P_n(x) = \sum_{\nu=0}^n d_{n\nu} \Phi_\nu(x) \ .
\end{equation}
 Besides the normalization factor $q_\nu$ let us also introduce, for
 later purposes, the integrals $p_\nu$ and $s_\nu$ by
\begin{eqnarray} \label{eq_11}
q_\nu &\equiv& \int_\epsilon^\lambda dx\, w(x)^2 \Phi_\nu(x)^2 \ ,
\nonumber \\[0.5em]
p_\nu &\equiv& \int_\epsilon^\lambda dx\, w(x)^2 \Phi_\nu(x)^2 x \ ,
\nonumber \\[0.5em]
s_\nu &\equiv& \int_\epsilon^\lambda dx\, w(x)^2 x^\nu \ .
\end{eqnarray}

 It can be easily shown that the expansion coefficients $d_{n\nu}$
 minimizing $\delta_n$ are independent of $n$ and are given by
\begin{equation} \label{eq_12}
d_{n\nu} \equiv d_\nu = \frac{b_\nu}{q_\nu} \ ,
\end{equation}
 where
\begin{equation} \label{eq_13}
b_\nu \equiv \int_\epsilon^\lambda dx\, w(x)^2 f(x) \Phi_\nu(x) \ .
\end{equation}
 The minimal value of $\delta_n^2$ is
\begin{equation} \label{eq_14}
\delta_n^2 = 1 - N_{\epsilon,\lambda}^{-1}
\sum_{\nu=0}^n d_\nu b_\nu \ .
\end{equation}

 Rescaling the variable $x$ by $x^\prime = \rho x$ allows for
 considering only standard intervals, say $[\epsilon/\lambda, 1]$.
 The scaling properties of the optimized polynomials can be easily
 obtained from the definitions.
 Let us now again consider the simple function $f(x)=x^{-\alpha}$ and
 relative deviation with $w(x)=x^\alpha$ when the rescaling relations
 are:
\begin{eqnarray} \label{eq_15}
\delta_n^2(\alpha;\epsilon\rho,\lambda\rho)
&=& \delta_n^2(\alpha;\epsilon,\lambda) \ ,
\nonumber \\[0.7em]
P_n(\alpha;\epsilon\rho,\lambda\rho;x) &=& \rho^{-\alpha}
P_n(\alpha;\epsilon,\lambda;x/\rho) \ ,
\nonumber \\[0.7em]
c_{n\nu}(\alpha;\epsilon\rho,\lambda\rho)
&=& \rho^{\nu-n-\alpha} c_{n\nu}(\alpha;\epsilon,\lambda) \ .
\end{eqnarray} 

 In applications to multi-bosonic algorithms for fermions the
 decomposition of the optimized polynomials as a product of root-factors
 is needed.
 This can be written as
\begin{equation} \label{eq_16}
P_n(\alpha;\epsilon,\lambda;x) =
c_{n0}(\alpha;\epsilon,\lambda) \prod_{j=1}^n 
[x-r_{nj}(\alpha;\epsilon,\lambda)] \ . 
\end{equation} 
 The rescaling properties here are:
\begin{eqnarray} \label{eq_17}
c_{n0}(\alpha;\epsilon\rho,\lambda\rho)
&=& \rho^{-n-\alpha}\, c_{n0}(\alpha;\epsilon,\lambda) \ ,
\nonumber \\[0.7em]
r_{nj}(\alpha;\epsilon\rho,\lambda\rho)
&=& \rho\, r_{nj}(\alpha;\epsilon,\lambda) \ .
\end{eqnarray} 
 The root-factorized form (\ref{eq_16}) can also be used for the
 numerical evaluation of the polynomials with matrix arguments if a
 suitable optimization of the ordering of roots is
 performed~\cite{POLYNOM}.

 The above orthogonal polynomials satisfy three-term recurrence
 relations which are very useful for numerical evaluation.
 In fact, at large $n$ the recursive evaluation of the polynomials is
 numerically more stable than the evaluation with root factors.
 For general $f(x)$ and $w(x)$, the first two ortogonal polynomials
 with $\mu=0,1$ are given by
\begin{equation} \label{eq_18}
\Phi_0(x) = 1 \ , \hspace{2em}
\Phi_1(x) = x - \frac{s_1}{s_0} \ .
\end{equation}
 The higher order polynomials $\Phi_\mu(x)$ for $\mu=2,3,\ldots$ can be
 obtained from the recurrence relation
\begin{equation} \label{eq_19}
\Phi_{\mu+1}(x) = (x+\beta_\mu)\Phi_\mu(x) +
\gamma_{\mu-1}\Phi_{\mu-1}(x) \ ,
\hspace{2em} (\mu=1,2,\ldots) \ ,
\end{equation}
 where the recurrence coefficients are given by
\begin{equation} \label{eq_20}
\beta_\mu = -\frac{p_\mu}{q_\mu} \ , \hspace{3em}
\gamma_{\mu-1} = -\frac{q_\mu}{q_{\mu-1}} \ .
\end{equation}
 Defining the polynomial coefficients $f_{\mu\nu}\; (0\leq\nu\leq\mu)$
 by
\begin{equation} \label{eq_21}
\Phi_\mu(x) = \sum_{\nu=0}^\mu f_{\mu\nu}x^{\mu-\nu}
\end{equation}
 the above recurrence relations imply the normalization convention
\begin{equation} \label{eq_22}
f_{\mu 0} = 1 \ , \hspace{3em} (\mu=0,1,2,\ldots) \ .
\end{equation}

 The rescaling relations for the orthogonal polynomials easily follow
 from the definitions.
 For the simple function $f(x)=x^{-\alpha}$ and relative deviation with
 $w(x)=x^\alpha$ we have
\begin{equation} \label{eq_23}
\Phi_\mu(\alpha;\rho\epsilon,\rho\lambda;x) = \rho^\mu\,
\Phi_\mu(\alpha;\epsilon,\lambda;x/\rho) \ .
\end{equation}
 For the quantities introduced in (\ref{eq_11}) this implies
\begin{eqnarray} \label{eq_24}
q_\nu(\alpha;\rho\epsilon,\rho\lambda) &=&
\rho^{2\alpha+1+2\nu}\, q_\nu(\alpha;\epsilon,\lambda) \ ,
\nonumber \\[0.7em]
p_\nu(\alpha;\rho\epsilon,\rho\lambda) &=&
\rho^{2\alpha+2+2\nu}\, p_\nu(\alpha;\epsilon,\lambda) \ ,
\nonumber \\[0.7em]
s_\nu(\alpha;\rho\epsilon,\rho\lambda) &=&
\rho^{2\alpha+1+\nu}\, s_\nu(\alpha;\epsilon,\lambda) \ .
\end{eqnarray}
 For the expansion coefficients defined in (\ref{eq_12})-(\ref{eq_13})
 one obtains
\begin{eqnarray} \label{eq_25}
b_\nu(\alpha;\rho\epsilon,\rho\lambda) &=&
\rho^{\alpha+1+\nu}\, b_\nu(\alpha;\epsilon,\lambda) \ ,
\nonumber \\[0.7em]
d_\nu(\alpha;\rho\epsilon,\rho\lambda) &=&
\rho^{-\alpha-\nu}\, d_\nu(\alpha;\epsilon,\lambda) \ ,
\end{eqnarray}
 and the recurrence coefficients in (\ref{eq_19})-(\ref{eq_20}) satisfy
\begin{eqnarray} \label{eq_26}
\beta_\mu(\alpha;\rho\epsilon,\rho\lambda) &=&
\rho\, \beta_\mu(\alpha;\epsilon,\lambda) \ ,
\nonumber \\[0.7em]
\gamma_{\mu-1}(\alpha;\rho\epsilon,\rho\lambda) &=&
\rho^2\, \gamma_{\mu-1}(\alpha;\epsilon,\lambda) \ .
\end{eqnarray}

 For general intervals $[\epsilon,\lambda]$ and/or functions
 $f(x)=x^{-\alpha}\bar{P}(x)$ the orthogonal polynomials and expansion
 coefficients have to be determined numerically.
 In some special cases, however, the polynomials can be related to
 some well know ones.
 An example is the weight factor
\begin{equation} \label{eq_27}
w^{(\rho,\sigma)}(x)^2 = (x-\epsilon)^\rho (\lambda-x)^\sigma \ .
\end{equation}
 Taking, for instance, $\rho=2\alpha,\; \sigma=0$ this weight is similar
 to the one for relative deviation from the function $f(x)=x^{-\alpha}$,
 which would be just $x^{2\alpha}$.
 In fact, for $\epsilon=0$ these are exactly the same and for small
 $\epsilon$ the difference is negligible.
 The corresponding orthogonal polynomials are simply related to the
 Jacobi polynomials \cite{DESYMUN2}, namely
\begin{equation} \label{eq_28}
\Phi^{(\rho,\sigma)}_\nu(x) = (\lambda-\epsilon)^\nu \nu!\,
\frac{\Gamma(\rho+\sigma+\nu+1)}{\Gamma(\rho+\sigma+2\nu+1)}
P^{(\sigma,\rho)}_\nu
\left(\frac{2x-\lambda-\epsilon}{\lambda-\epsilon}\right) \ .
\end{equation}
 Comparing different approximations with different $(\rho,\sigma)$
 the best choice is usually $\rho=2\alpha,\; \sigma=0$ which
 corresponds to optimizing the relative deviation (see the appendix of
 \cite{DESYMUN2}).

\begin{figure}[ht]
\vspace{-1.5cm}
\begin{center}
\epsfig{file=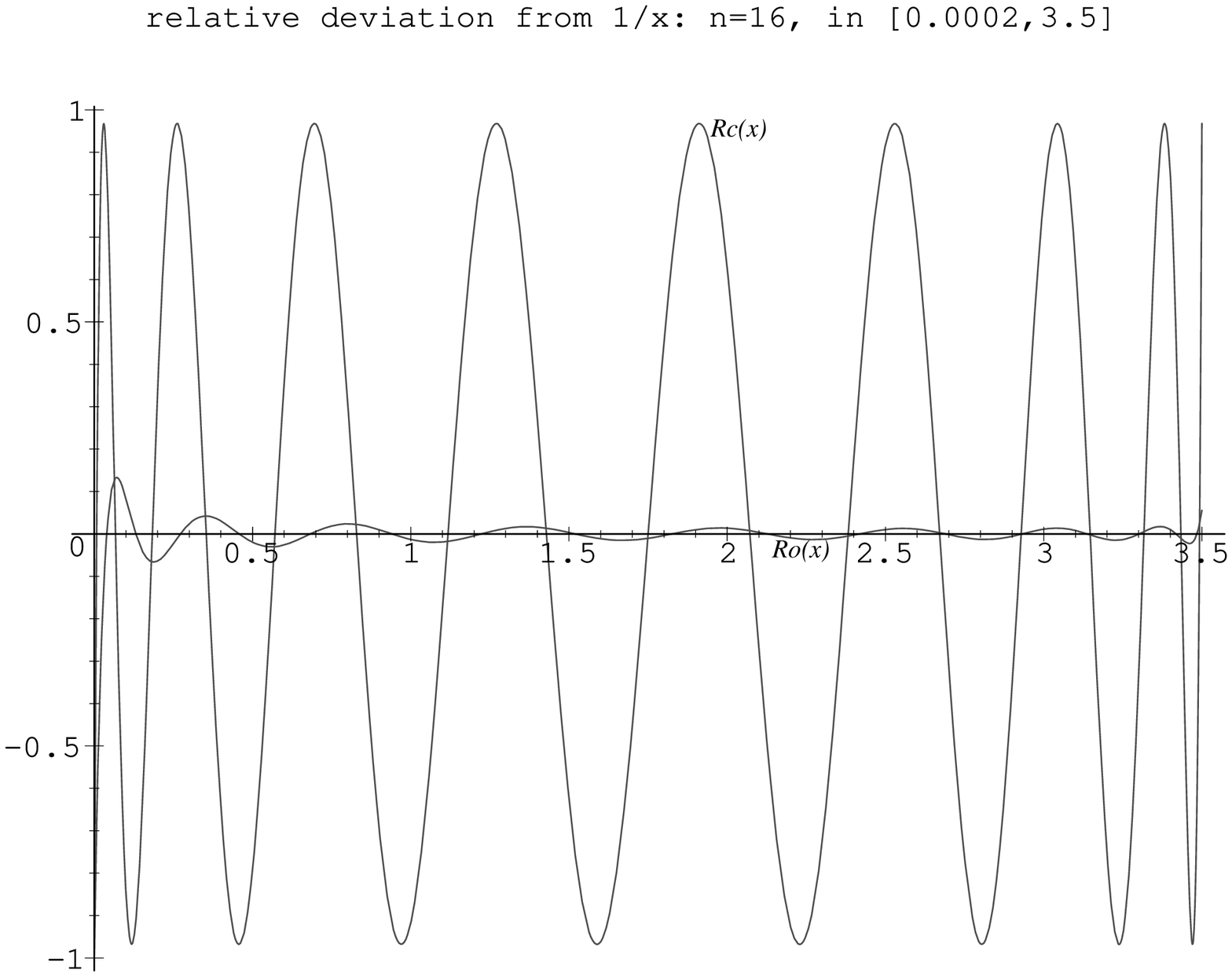,
        width=11.7cm,height=11.7cm,
        angle=270}
\end{center}
\vspace{-2.7cm}
\begin{flushleft}
\epsfig{file=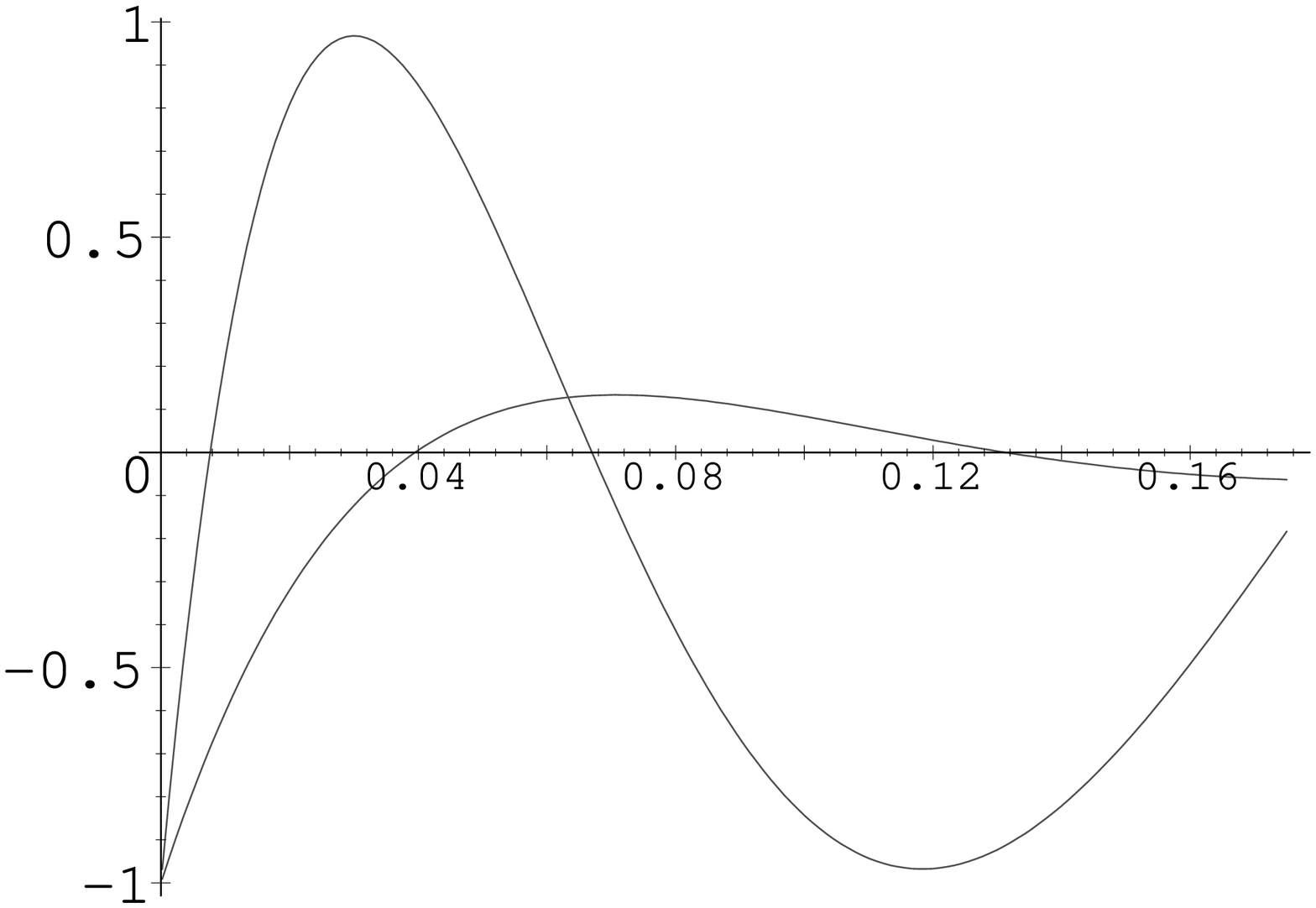,
        width=6.5cm,height=6.5cm,
        angle=270}
\end{flushleft}
\vspace{-7.2cm}
\begin{flushright}
\epsfig{file=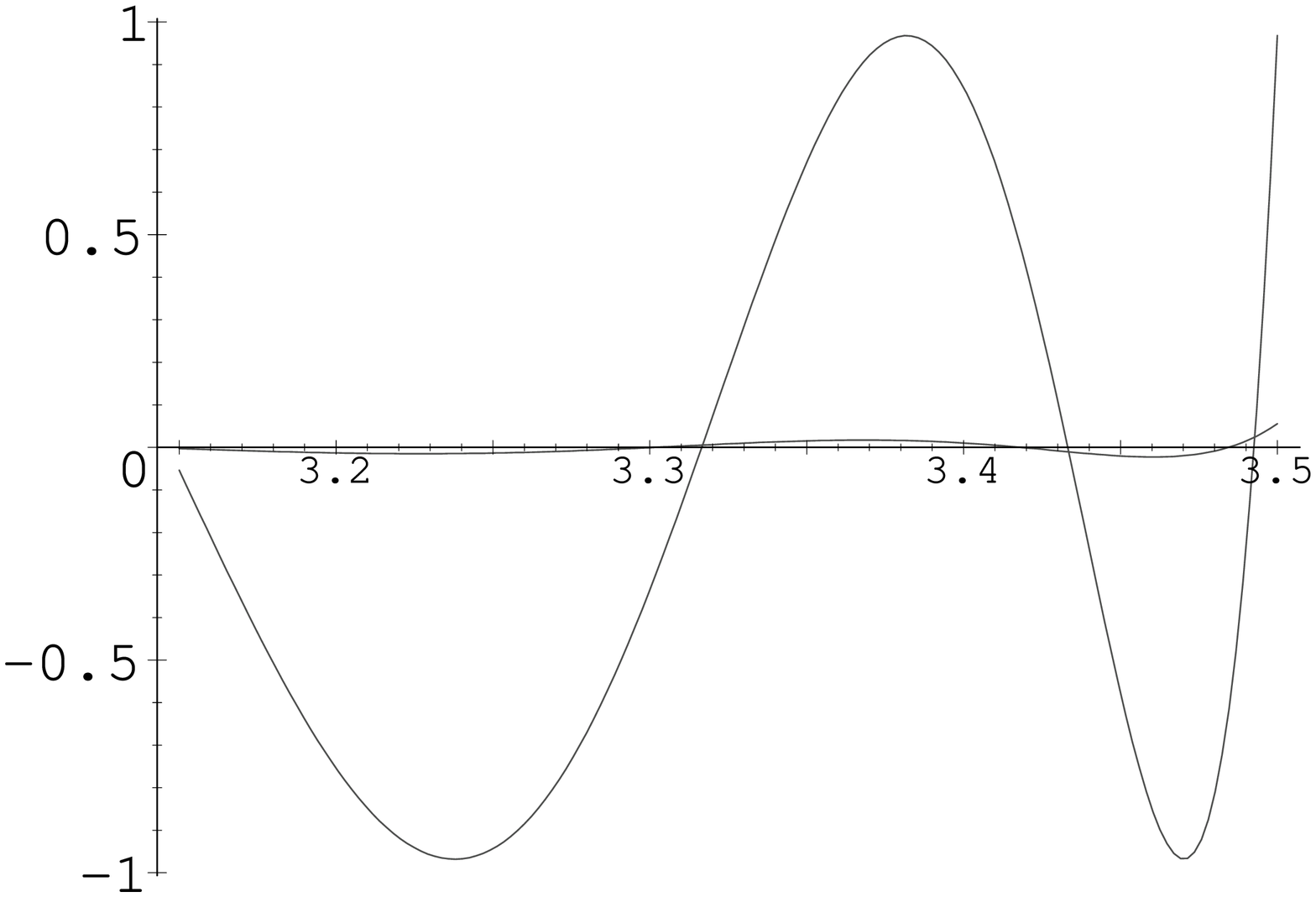,
        width=6.5cm,height=6.5cm,
        angle=270}
\end{flushright}
\vspace{-0.7cm}
\begin{center}
\caption{\label{ChebyFig}
 The comparison of the relative deviation $R(x) \equiv xP(x)-1$ for the
 Chebyshev polynomial $(Rc(x))$ and the quadratically optimized 
 polynomial $(Ro(x))$.
 In the lower part the two ends of the interval are zoomed.}
\end{center}
\end{figure}

 For large condition numbers $\lambda/\epsilon$ least-squares
 optimization is much better than the Chebyshev approximation used for
 the approximation of $x^{-1}$ in \cite{MULTIBOSON}.
 The Chebyshev approximation is minimizing the maximum of the relative
 deviation
\begin{equation} \label{eq_29}
R(x) \equiv xP(x)-1 \ .
\end{equation}
 For the deviation norm
\begin{equation} \label{eq_30}
\delta_{max} \equiv \max_{x \in [\epsilon,\lambda]} |R(x)|
\end{equation}
 the least-squares approximation is slightly worse than the
 Chebyshev approximation.
 An example is shown by fig.~\ref{ChebyFig}.
 In the left lower corner the Chebyshev approximation has
 $Rc(0.0002)=-0.968$ compared to $Ro(0.0002)=-0.991$ for the
 least-squares optimization.
 For smaller condition numbers the Chebyshev approximation is not as
 bad as is shown by fig.~\ref{ChebyFig}.
 Nevertheless, in QCD simulations in sufficiently large volumes the
 condition number is of the order of the light quark mass squared in
 lattice units which can be as large as ${\cal O}(10^6-10^7)$.
 
 Figure ~\ref{ChebyFig} also shows that the least-squares optimization
 is quite good in the minimax norm in (\ref{eq_30}), too.
 It can be proven that
\begin{equation} \label{eq_31}
|Ro(\epsilon)| = \max_{x \in [\epsilon,\lambda]}\, |Ro(x)|
\end{equation}
 hence the minimax norm can also be easily obtained from
\begin{equation} \label{eq_32}
\delta_{max}^{(o)} = \max_{x \in [\epsilon,\lambda]}\,|Ro(x)| =
 |Ro(\epsilon)| \ .
\end{equation}
 Therefore the least squares-optimization is also well suited for
 controlling the minimax norm, if for some reason it is required.

\begin{figure}[ht]
\begin{center}
\epsfig{file=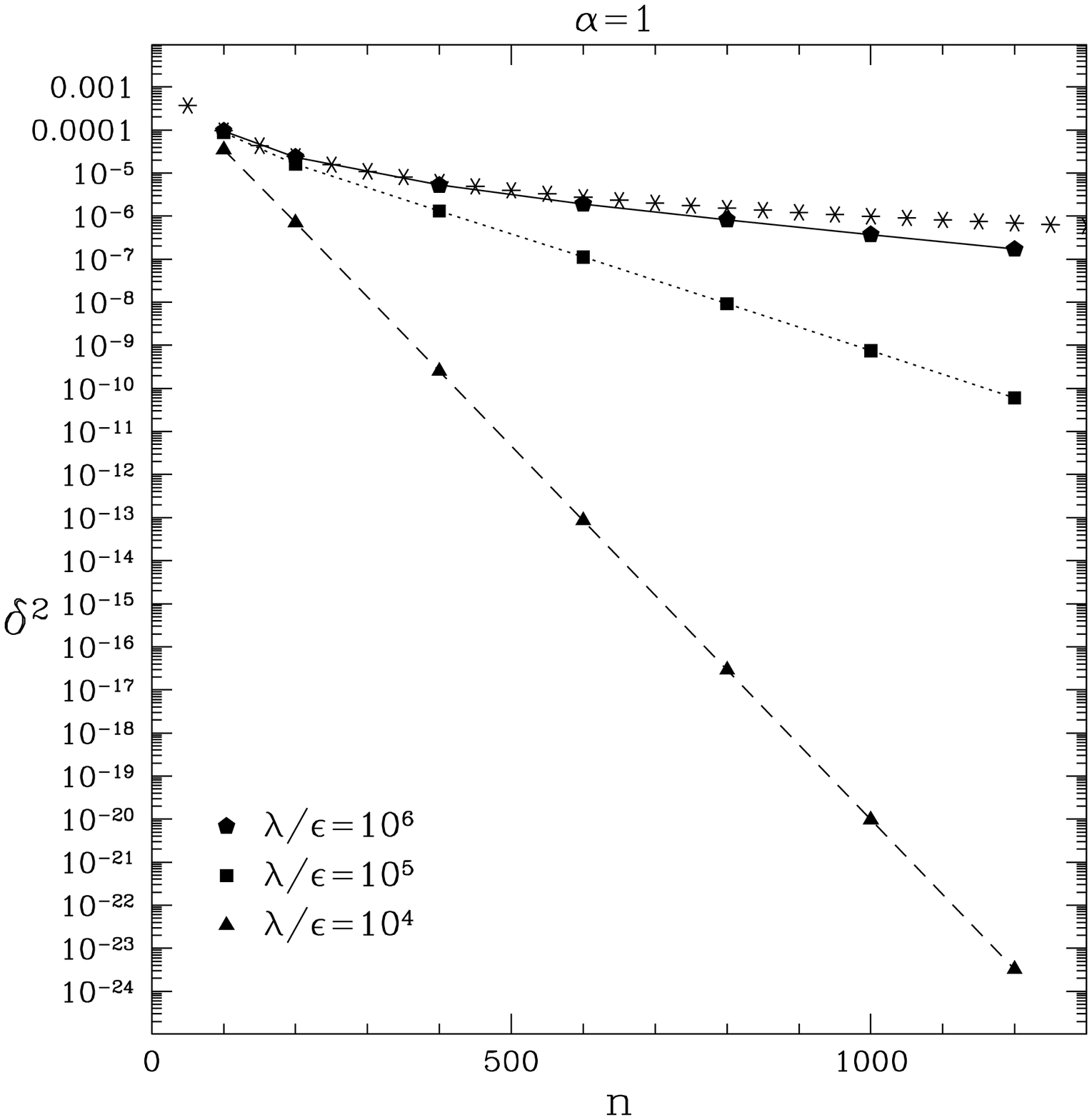,
        width=11.7cm,height=11.7cm,
        angle=0}
\end{center}
\vspace{-0.7cm}
\begin{center}
\caption{\label{Alpha=1Fig}
 The (squared) deviation norm $\delta^2$ of the polynomial
 approximations of $x^{-1}$ as function of the order for different
 values of $\lambda/\epsilon$.
 The asterisks show the $\epsilon/\lambda \to 0$ limit.}
\end{center}
\end{figure}

\begin{figure}[ht]
\begin{center}
\epsfig{file=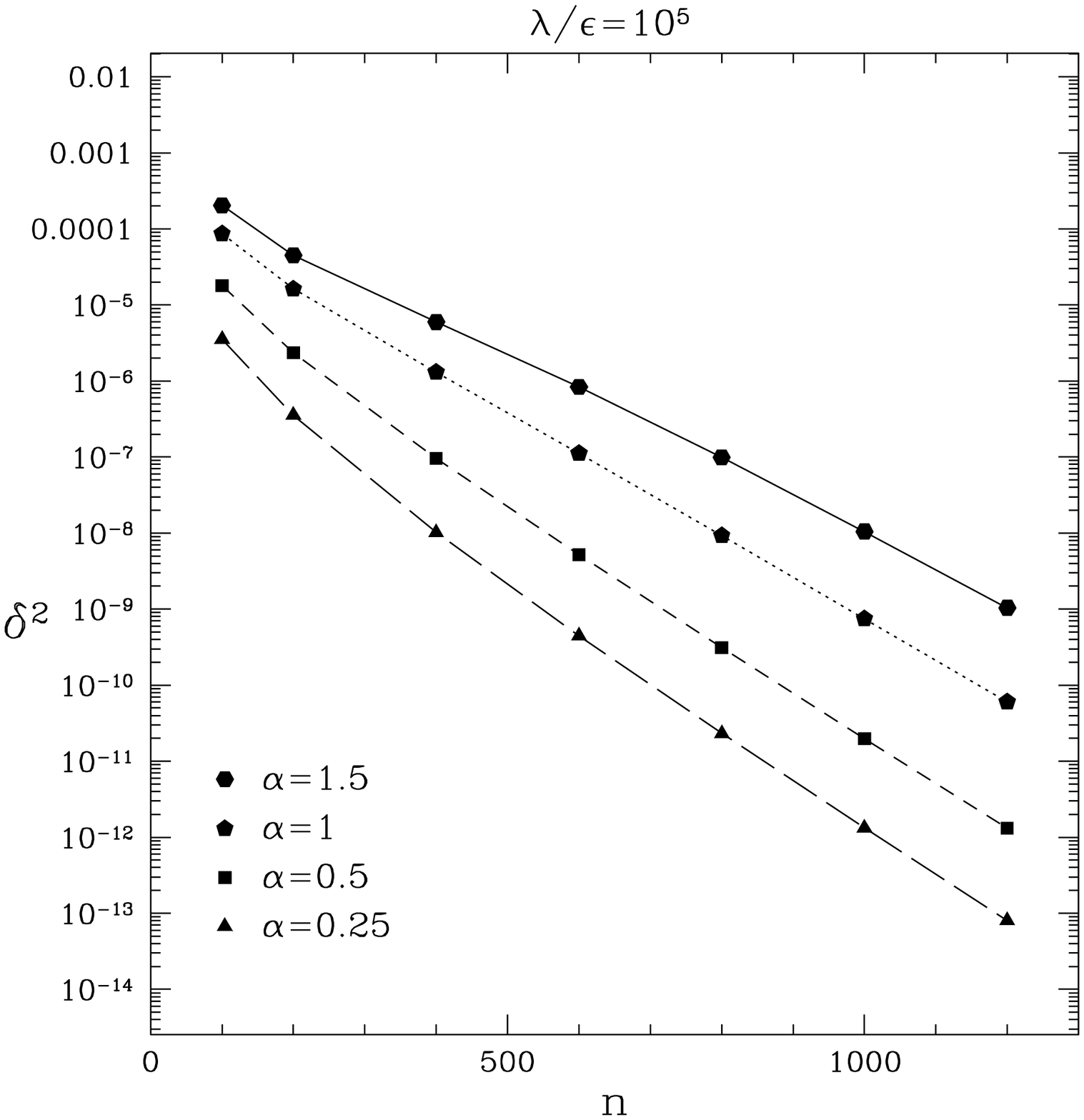,
        width=11.7cm,height=11.7cm,
        angle=0}
\end{center}
\vspace{-0.7cm}
\begin{center}
\caption{\label{10^5Fig}
 The (squared) deviation norm $\delta^2$ of the polynomial
 approximations of $x^{-\alpha}$ as function of the order at
 $\lambda/\epsilon=10^5$ for different values of $\alpha$.}
\end{center}
\end{figure}

 In QCD simulations the inverse power to be approximated ($\alpha$) is
 related to the number of Dirac fermion flavours: $\alpha=N_f/2$.
 If only $u$- and $d$-quarks are considered we have $N_f=2$ and the
 function to be approximated is $x^{-1}$.
 The dependence of the (squared) least-squares norm in (\ref{eq_01}) on
 the polynomial order $n$ is shown by fig.~\ref{Alpha=1Fig} for
 different values of the condition number $\lambda/\epsilon$.
 The dependence on $\alpha=N_f/2$ is illustrated by fig.~\ref{10^5Fig}.

 Another possible application of least-squares optimized polynomials
 is the numerical evaluation of the zero mass lattice action proposed by
 Neuberger~\cite{NEUBERGER}.
 If one takes, for instance, the weight factor in (\ref{eq_27})
 corresponding to the relative deviation, then the function
 $x^{-1/2}$ has to be expanded in the Jacobi polynomials $P^{(1,0)}$.

\section{Recurrence scheme}

 The expansion in orthogonal polynomials is very useful because it
 allows for a numerically stable evaluation of the least-squares
 optimized polynomials by the recurrence relation (\ref{eq_19}).
 The orthogonal polynomials themselves can also be determined
 recursively.

 A recurrence scheme for obtaining the recurrence coefficients
 $\beta_\mu,\gamma_{\mu-1}$ and expansion coefficients
 $d_\nu= b_\nu/q_\nu$ has been given in \cite{DESYMUN1,MOLDYN}.
 In order to obtain $q_\nu,p_\nu$ contained in (\ref{eq_20}) one can
 use the relations
\begin{equation} \label{eq_33}
q_\mu = \sum_{\nu=0}^\mu f_{\mu\nu} s_{2\mu-\nu} \ ,
\hspace{2em}
p_\mu = \sum_{\nu=0}^\mu f_{\mu\nu}
\left( s_{2\mu+1-\nu} +f_{\mu 1}s_{2\mu-\nu} \right) \ .
\end{equation}
 The coefficients themselves can be calculated from $f_{11}=-s_1/s_0$
 and (\ref{eq_19}) which gives
\begin{eqnarray}\label{eq_34}
f_{\mu+1,1} & = & f_{\mu,1} + \beta_\mu \ ,             \nonumber \\
f_{\mu+1,2} & = & f_{\mu,2} + \beta_\mu f_{\mu,1} +
\gamma_{\mu-1} \ ,                                      \nonumber \\
f_{\mu+1,3} & = & f_{\mu,3} + \beta_\mu f_{\mu,2} +
\gamma_{\mu-1} f_{\mu-1,1} \ ,                          \nonumber \\
            & \ldots &                                  \nonumber \\
f_{\mu+1,\mu} & = & f_{\mu,\mu} + \beta_\mu f_{\mu,\mu-1} +
\gamma_{\mu-1} f_{\mu-1,\mu-2} \ ,                      \nonumber \\
f_{\mu+1,\mu+1} & = & \beta_\mu f_{\mu,\mu} +
\gamma_{\mu-1} f_{\mu-1,\mu-1} \ .
\end{eqnarray}
 The orthogonal polynomial and recurrence coefficients are recursively
 determined by (\ref{eq_20}) and (\ref{eq_33})-(\ref{eq_34}).
 The expansion coefficients for the optimized polynomial $P_n(x)$
 can be obtained from
\begin{equation} \label{eq_35}
b_\mu = \sum_{\nu=0}^\mu f_{\mu\nu}
\int_\epsilon^\lambda dx\, w(x)^2 f(x) x^{\mu-\nu} \ .
\end{equation}
 The ingredients needed for this recursion are the basic integrals
 $s_\nu$ defined in (\ref{eq_11}) and
\begin{equation} \label{eq_36}
t_\nu \equiv \int_\epsilon^\lambda dx\, w(x)^2 f(x) x^\nu \ .
\end{equation}

 The recurrence scheme based on the coefficients of the orthogonal
 polynomials $f_{\mu\nu}$ in (\ref{eq_34}) is not optimal for large
 orders $n$, neither for arithmetics nor for storage requirements.
 A better scheme can be built up on the basis of the integrals
\begin{eqnarray} \label{eq_37}
& & r_{\mu\nu} \equiv
\int_\epsilon^\lambda dx\, w(x)^2 \Phi_\mu(x) x^\nu \ ,
\nonumber \\[0.5em]
& & \mu=0,1,\ldots,n;
\hspace{1.5em} \nu=\mu,\mu+1,\ldots,2n-\mu \ .
\end{eqnarray}
 The recurrence coefficients $\beta_\mu,\;\gamma_{\mu-1}$ can be
 expressed from
\begin{equation} \label{eq_38}
q_\mu = r_{\mu\mu} \ ,\hspace{2em} 
p_\mu=r_{\mu,\mu+1} + f_{\mu 1}r_{\mu\mu}
\end{equation}
 and eq.~(\ref{eq_20}) as
\begin{equation} \label{eq_39}
\beta_\mu = - f_{\mu 1} - \frac{r_{\mu,\mu+1}}{r_{\mu\mu}} \ ,
\hspace{2em}
\gamma_{\mu-1} = - \frac{r_{\mu\mu}}{r_{\mu-1,\mu-1}} \ .
\end{equation}
 It follows from the definition that
\begin{eqnarray} \label{eq_40}
r_{0\nu} & = & \int_\epsilon^\lambda dx\, w(x)^2 x^\nu = s_\nu \ ,
\nonumber \\[0.5em]
r_{1\nu} & = & \int_\epsilon^\lambda dx\, w(x)^2 
(x^{\nu+1} + f_{11}x^\nu) = s_{\nu+1} + f_{11}s_\nu \ .
\end{eqnarray}
 The recurrence relation (\ref{eq_19}) for the orthogonal polynomials
 implies
\begin{equation} \label{eq_41}
r_{\mu+1,\nu} = r_{\mu,\nu+1} + \beta_\mu r_{\mu\nu}
+ \gamma_{\mu-1}r_{\mu-1,\nu} \ .
\end{equation}
 This has to be supplemented by
\begin{equation} \label{eq_42}
f_{11} = -\frac{s_1}{s_0}
\end{equation}
 and by the first equation from (\ref{eq_34}):
\begin{equation} \label{eq_43}
f_{\mu+1,1} = f_{\mu,1} + \beta_\mu \ .
\end{equation}
 Eqs.~(\ref{eq_39})-(\ref{eq_43}) define a complete recurrence scheme
 for determining the orthogonal polynomials $\Phi_\mu(x)$.
 The moments $s_\nu$ of the integration measure defined in (\ref{eq_11})
 serve as the basic input in this scheme.

 The integrals $b_\nu$ in (\ref{eq_13}), which are necessary for the
 expansion coefficients $d_\nu$ in (\ref{eq_12}), can also be calculated
 in a similar scheme built up on the integrals
\begin{eqnarray} \label{eq_44}
& & b_{\mu\nu} \equiv
\int_\epsilon^\lambda dx\, w(x)^2 f(x) \Phi_\mu(x) x^\nu \ ,
\nonumber \\[0.5em]
& & \mu=0,1,\ldots,n;
\hspace{1.5em} \nu=0,1,\ldots,n-\mu \ .
\end{eqnarray}
 The relations corresponding to (\ref{eq_40})-(\ref{eq_41}) are now
\begin{eqnarray} \label{eq_45}
& & b_{0\nu} = \int_\epsilon^\lambda dx\, w(x)^2 f(x) x^\nu = t_\nu \ ,
\nonumber \\[0.5em]
& & b_{1\nu} = \int_\epsilon^\lambda dx\, w(x)^2 f(x) 
(x^{\nu+1} + f_{11}x^\nu) = t_{\nu+1} + f_{11}t_\nu \ ,
\nonumber \\[0.5em]
& & b_{\mu+1,\nu} = b_{\mu,\nu+1} + \beta_\mu b_{\mu\nu}
+ \gamma_{\mu-1}b_{\mu-1,\nu} \ .
\end{eqnarray}
 The only difference compared to (\ref{eq_40})-(\ref{eq_41}) is that
 the moments of $w(x)^2$ are now replaced by the ones of $w(x)^2f(x)$.

 It is interesting to collect the quantities which have to be stored in
 order that the recurrence can be resumed.
 This is useful if after stopping the iterations, for some reason, the
 recurrence will be restarted.
 Let us assume that the quantities $q_\nu\;(\nu=0,\ldots,n)$,
 $b_\nu\;(\nu=0,\ldots,n)$, $\beta_\nu\;(\nu=1,\ldots,n-1)$ and
 $\gamma_\nu\;(\nu=0,\ldots,n-2)$ are already known and one wants to
 resume the recurrence in order to calculate these quantities for
 higher indices.
 For this it is enough to know the values of
\begin{eqnarray} \label{eq_46}
& & f_{n-1,1} \ ,\;\; r_{n-1,n-1} \ ,
\nonumber \\[0.5em]
& & R^{(0)}_{0 \ldots n} \equiv 
(r_{0,2n+1},r_{1,2n},\ldots,r_{n,n+1}) \ ,
\nonumber \\[0.5em]
& & R^{(1)}_{0 \ldots n} \equiv 
(r_{0,2n},r_{1,2n-1},\ldots,r_{n,n}) \ ,
\nonumber \\[0.5em]
& & B^{(1)}_{0 \ldots n} \equiv 
(b_{0,n},b_{1,n-1},\ldots,b_{n,0}) \ ,
\nonumber \\[0.5em]
& & B^{(2)}_{0 \ldots n-1} \equiv 
(b_{0,n-1},b_{1,n-2},\ldots,b_{n-1,0}) \ .
\end{eqnarray}
 This shows that for maintaining a {\em resumable recurrence} it is
 enough to store a set of quantities linearly increasing in $n$.

 An interesting question is the increase of computational load as
 a function of the highest required order $n$.
 At the first sight this seems to go just like $n^2$, which is
 surprising because, as eq.~(\ref{eq_06}) shows, finding the minimum
 requires the inversion of an $n \otimes n$ matrix.
 However, numerical experience shows that the number of required digits
 for obtaining a precise result does also increase linearly with $n$.
 This is due to the linearly increasing logarithmic range of
 eigenvalues, as illustrated by (\ref{eq_08}).
 Using, for instance, Maple V for the arbitrary precision arithmetic,
 the computation slows down by another factor going roughly as
 (but somewhat slower than) $n^2$.
 Therefore, the total slowing down in $n$ is proportional to $n^4$.
 For the same reason the storage requirements increase by $n^2$.

\section{A convenient choice for TSMB}

 In the TSMB algorithm for Monte Carlo simulations of fermionic
 theories, besides the simple function $x^{-\alpha}$, also the function
 $x^{-\alpha}/\bar{P}(x)$ has to be approximated.
 Here $\bar{P}(x)$ is typically a lower order approximation to
 $x^{-\alpha}$.
 In this case, if one chooses to optimize the relative deviation,
 the basic integrals defined in (\ref{eq_11}) and (\ref{eq_36}) are,
 respectively,
\begin{eqnarray} \label{eq_47}
s_\nu & = & \int_\epsilon^\lambda dx\, \bar{P}(x)^2 x^{2\alpha+\nu} \ ,
\nonumber \\[0.5em]
t_\nu & = & \int_\epsilon^\lambda dx\, \bar{P}(x) x^{\alpha+\nu} \ .
\end{eqnarray}
 It is obvious that, if the recurrence coefficients for the expansion
 of the polynomial $\bar{P}(x)$ in orthogonal polynomials are known,
 the recursion scheme can also be used for the evaluation of $s_\nu$ and
 $t_\nu$.

 Another observation is that the integrals in (\ref{eq_47}) can be
 simplifyed if, instead of choosing the weight factor
 $w(x)^2=\bar{P}(x)^2 x^{2\alpha}$, one takes
\begin{equation} \label{eq_48}
w(x)^2 = \bar{P}(x) x^\alpha \ ,
\end{equation}
 which leads to
\begin{eqnarray} \label{eq_49}
s_\nu & = & \int_\epsilon^\lambda dx\, \bar{P}(x) x^{\alpha+\nu} \ ,
\nonumber \\[0.5em]
t_\nu & = & \int_\epsilon^\lambda dx\, x^\nu \ .
\end{eqnarray}
 Since $\bar{P}(x)$ is an approximation to $x^{-\alpha}$, the function
 $f(x) \equiv x^{-\alpha}/\bar{P}(x)$ is close to one and the difference
 between $f(x)^{-2}$ and $f(x)^{-1}$ is small.
 Therefore the least-squares optimized approximations with the weights
 $w(x)^2=f(x)^{-2}$ and $w(x)^2=f(x)^{-1}$ are also similar.
 It turns out that the second choice is, in fact, a little bit better
 because the largest deviation from $x^{-\alpha}$ typically occurs
 at the lower end of the interval $x=\epsilon$ where $\bar{P}(x)$ is
 smaller than $x^{-\alpha}$.
 As a consequence, $\bar{P}(x) x^\alpha < 1$ and
 $\bar{P}(x) x^{\alpha} > \bar{P}(x)^2 x^{2\alpha}$.
 This means that choosing the weight factor in (\ref{eq_48}) is
 emphasising more the lower end of the interval where $\bar{P}(x)$ as an
 approximation of $x^{-\alpha}$ is worst.

 In summary: least-squares optimization is a flexible and powerful tool
 which can serve as a basis for applying the two-step multi-bosonic
 algorithm for Monte Carlo simulations of QCD and other similar
 theories.
 With the help of the recurrence scheme described in the previous
 section one can determine the necessary polynomial approximations to
 high enough orders.


\end{document}